\documentclass[aps,prb,twocolumn,groupedaddress,amsmath,letterpaper]{revtex4}

\usepackage{amsmath,amssymb,graphicx}
\usepackage{array}
\usepackage{slashed}

\begin{document}

\title{Electromagnetic Energy Momentum Tensor in a Spatially Dispersive Medium}

\author{Chris Fietz}
\email[Email: ]{fietz.chris@gmail.com}
\affiliation{Edinburg, Texas, USA}

\begin{abstract}
We derive a generalized Minkowski Energy Momentum Tensor for a monochromatic wave in a lossless medium exhibiting temporal and spatial dispersion.  The Energy Momentum Tensor is then related to familiar expressions for energy density and energy flux, as well as new expressions for momentum density and momentum flux.
\end{abstract}


\maketitle

\section{Motivation}
As interest in photonic crystals and metamaterials expands, it is becoming increasingly clear that these materials often exhibit strong spatial dispersion.  Understanding spatial dispersion and its consequences is essential to understanding these materials.  It has long been known~\cite{Brillouin_60} that temporal dispersion alters the expression for the electromagnetic energy density of a material.  Similarly, it is known that spatial dispersion alters the expression for the Poynting energy flux\cite{Agranovich_Energy,Landau_Energy}.  A simple generalization of these arguments shows that temporal and spatial dispersion affect the Energy Momentum Tensor (EM Tensor) as a whole.  Here we derive the entire EM Tensor for a monochromatic wave in a temporally and spatially dispersive medium with no loss.  We then show that for a freely propagating wave the various components of the EM Tensor are related by the group and phase velocities.  From the EM Tensor, we then derive more familiar expressions for the energy density, energy flux, momentum density, and momentum flux.  Finally, we introduce a small amount of loss to see how this effects energy-momentum conservation.

\section{Setup}

We define the electric field and magnetic flux density in terms of the electromagnetic 4-vector potential $\mathrm{A}_{\mu} = (\mathrm{A}_0,-\mathbf{\mathbf{A}})$

\begin{equation}\label{EB_def}
\mathbf{E} = -\nabla\mathrm{A}_0 - \dfrac{1}{c}\dfrac{\partial\mathbf{A}}{\partial t},
\ \ \ \ \ \ \ \ 
\mathbf{\mathbf{B}} = \nabla\times\mathbf{A}.
\end{equation} 

\noindent  The definition of $\mathbf{E}$ and $\mathbf{B}$ implies one half of the Maxwell equations


\begin{equation}\label{Max_1}
\nabla\cdot\mathbf{B}=0,
\ \ \ \ \ \ \ \ 
-\nabla\times\mathbf{E}-\dfrac{1}{c}\dfrac{\partial\mathbf{B}}{\partial t} = 0.
\end{equation} 

\noindent  The remaining Maxwell equations connect the potential fields to the source 4-vector $\mathrm{J}^{\mu}=(c\rho,\mathbf{J})$

\begin{equation}\label{Max_2}
\nabla\cdot\mathbf{D}=\rho,
\ \ \ \ \ \ \ \ 
\nabla\times\mathbf{H}-\dfrac{1}{c}\dfrac{\partial\mathbf{D}}{\partial t} = \dfrac{\mathbf{J}}{c}.
\end{equation} 

\noindent Here we have used Heaviside-Lorentz units~\cite{Jackson_Units}.  Eq.~(\ref{Max_2}) requires the use of the constitutive parameters

\begin{equation}\label{constitutive}
\begin{array}{rl}

\mathbf{D}(\omega,\mathbf{k}) = & \hat{p}(\omega,\mathbf{k})\cdot\mathbf{E}(\omega,\mathbf{k}) + \hat{l}(\omega,\mathbf{k})\cdot\mathbf{B}(\omega,\mathbf{k}), \\ \\

\mathbf{H}(\omega,\mathbf{k}) = & \hat{m}(\omega,\mathbf{k})\cdot\mathbf{E}(\omega,\mathbf{k}) + \hat{q}(\omega,\mathbf{k})\cdot\mathbf{B}(\omega,\mathbf{k}),

\end{array}
\end{equation}

\noindent  where $\hat{p}$ and $\hat{q}$ are $3\times 3$ spatial tensors while $\hat{l}$ and $\hat{m}$ are $3\times 3$ spatial pseudotensors.  The constitutive parameters are all functions of the frequency $\omega$, implying temporal dispersion, as well as the wavevector $\mathbf{k}$, implying spatial dispersion.  Defining the constitutive parameters as functions of frequency implies that the medium is not changing in time.  Similarly, defining the constitutive parameters as functions of the wavevector implies that the medium has infinitesimal translational symmetry (the medium does not change as a function of space).

We now define the Fourier transformation between real and reciprocal spacetime as

\begin{equation}
\mathrm{F}(x) = \displaystyle\int\dfrac{d^4k}{(2\pi)^2} \mathrm{F}(k)e^{\mathrm{i}k\cdot x},
\end{equation} 

\noindent where $k\cdot x = k_{\mu}x^{\mu}$ is the dot product between the wave 4-vector $k_{\mu}=(\omega/c,-\mathbf{k})$ and the position 4-vector $x^{\mu}=(ct,\mathbf{x})$.  $\mathrm{i}k_{\mu}$ is itself the reciprocal of the differential operator $\partial_{\mu}=(\frac{1}{c}\frac{\partial}{\partial t},\nabla)$.  Throughout this paper we assume the Minkowski metric $\eta_{\mu\nu}=(1,-1,-1,-1)$ for flat spacetime.

In this paper we use a combination of Greek and Latin indices.  As is normally the convention, Greek indices vary over the four spacetime dimensions $(0,1,2,3)$ while Latin indices vary over only the spatial dimensions $(1,2,3)$.  Anytime an index is repeated, a summation over the possible index values is implied.

Combining Eqs.~(\ref{EB_def},\ref{Max_2},\ref{constitutive}), we obtain an wave equation in reciprocal (frequency-wavevector) space

\begin{equation}\label{wave_eq}
\dfrac{1}{c}\mathrm{J}^{\alpha}(k) = D^{\alpha\beta}(k) \mathrm{A}_{\beta}(k). \\ \\
\end{equation}

\noindent  Here we have used the reciprocal differential operator

\begin{equation}\label{D_oper}
D^{\alpha\beta}(k) = k_{\mu}C^{\alpha\mu\nu\beta}k_{\nu},
\end{equation}

\noindent which is itself defined using the constitutive tensor~\cite{Kong_Constitutive}

\begin{equation}\label{C_con}
\begin{array}{rcl}
C^{0 i 0 j} & = & -p_{ij},
\\
C^{ijkl} & = & \epsilon_{ijm} \epsilon_{kln} q_{mn},
\\
C^{0kij} & = & \epsilon_{ijm} l_{km},
\\
C^{ij0k} & = & -\epsilon_{ijn} m_{nk}.
\end{array}
\end{equation}

\noindent  Here we have used the three dimensional fully antisymmetric Levi-Civita symbol using the convention $\epsilon_{123}=+1$.  The constitutive tensor is skew symmetric, which means that it changes sign upon the interchange of either the first pair or last pair of indices, or $C^{\alpha\mu\nu\beta}=-C^{\alpha\mu\beta\nu}=-C^{\mu\alpha\nu\beta}=C^{\mu\alpha\beta\nu}$.  Additionally, for a lossless medium, the constitutive tensor has the property $C^{\alpha\mu\nu\beta}=(C^{\nu\beta\alpha\mu})^*$, where $*$ implies complex conjugation.  Combined with Eq.~(\ref{D_oper}), this implies that for a lossless medium the reciprocal differential operator is Hermitian or $D^{\alpha\beta}=(D^{\beta\alpha})^*$.

One should not focus too much on the form of the reciprocal differential operator $\mathrm{D}^{\alpha\beta}$.  We present it here simply for completeness.  Most of our results are general and independent of the exact form of $\mathrm{D}^{\alpha\beta}$.

\section{Derivation}

Central to our derivation is restriction of the various fields to being approximately monochromatic.  The electromagnetic 4-vector is represented as 

\begin{equation}
\mathrm{A}_{\mu}(x) = \dfrac{\mathrm{A}_{\mu}^{\prime}(x)e^{\mathrm{i}k\cdot x} + c.c.}{2}.
\end{equation}

\noindent  Here $\mathrm{A}_{\mu}^{\prime}(x)$ is a slowly varying (in time and space) complex valued amplitude and $c.c.$ represents the complex conjugate terms.  $k_{\mu}$ is the real valued frequency and wavevector for the carrier wave.  Throughout this paper all \textit{primed} fields represent complex valued amplitudes that are slowly varying in real spacetime.  The amplitude $\mathrm{A}_{\mu}^{\prime}(x)$ can itself be expressed in reciprocal spacetime

\begin{equation}
\mathrm{A}_{\mu}^{\prime}(q) = \displaystyle\int\dfrac{d^4x}{(2\pi)^2} \mathrm{A}_{\mu}^{\prime}(x)e^{-\mathrm{i}q\cdot x}.
\end{equation}

\noindent  Since the amplitude $\mathrm{A}^{\prime}(x)$ is a slowly varying in real spacetime, the reciprocal amplitude $\mathrm{A}^{\prime}(q)$ has a very narrow bandwidth in reciprocal spacetime.

We now express the complex valued source 4-vector amplitude $\mathrm{J}^{\prime\alpha}$ in terms of the reciprocal differential operator $D^{\alpha\beta}$ and the 4-vector potential amplitude $\mathrm{A}_{\beta}^{\prime}$

\begin{equation}\label{J_prime}
\begin{array}{rl}

\dfrac{1}{c}\mathrm{J}^{\prime\alpha}(x) = & \displaystyle\int\dfrac{d^4q}{(2\pi)^2} D^{\alpha\beta}(k+q) \mathrm{A}_{\beta}^{\prime}(q) e^{\mathrm{i}q\cdot x} \\ \\

 \approx & \displaystyle\int\dfrac{d^4q}{(2\pi)^2} \left( D^{\alpha\beta}(k) + q_{\eta}\dfrac{\partial D^{\alpha\beta}(k)}{\partial k_{\eta}} \right) \mathrm{A}_{\beta}^{\prime}(q) e^{\mathrm{i}q\cdot x} \\ \\

 = & D^{\alpha\beta}(k)\mathrm{A}_{\beta}^{\prime}(x) -\mathrm{i}\dfrac{\partial D^{\alpha\beta}(k)}{\partial k_{\eta}}\partial_{\eta}\mathrm{A}_{\beta}^{\prime}(x).

\end{array}
\end{equation}

\noindent Here we have expanded $D^{\alpha\beta}(k+q)$ in $q$.  This maneuver is similar in principle to the Poynting Flux derivations in Refs. \onlinecite{Agranovich_Energy,Landau_Energy,Kamenetskii_96}.  The expansion to only first order is justified by the very narrow bandwidth of the reciprocal amplitude $\mathrm{A}^{\prime}(q)$.

We connect the source 4-vector to the EM Tensor with the time averaged force density 4-vector

\begin{equation}\label{force}
\mathrm{F}^{\mu} = \langle(\partial^{\mu}\mathrm{A}^{\nu}-\partial^{\nu}\mathrm{A}^{\mu}) \mathrm{J}_{\nu}/c\rangle = 
\left(\!\!\begin{array}{c}
\langle\mathbf{E}\cdot\mathbf{J}/c\rangle \\ \langle\rho\mathbf{E}+\dfrac{1}{c}\mathbf{J}\times\mathbf{B}\rangle
\end{array}\!\!\right),
\end{equation}

\noindent  where time averaging is indicated by $\langle \rangle$.  By combining Eqs.~(\ref{J_prime}) and (\ref{force}) we obtain an expression for the time averaged force 4-vector in terms of the slowly varying amplitude $\mathrm{A}_{\mu}^{\prime}(x)$

\begin{widetext}
\begin{equation}\label{f_av}
\begin{array}{rl}

\mathrm{F}^{\nu} = & \dfrac{1}{4} \Bigl( (\mathrm{i}k^{\nu}\mathrm{A}^{\prime\eta}+\partial^{\nu}\mathrm{A}^{\prime\eta} - \mathrm{i}k^{\eta}\mathrm{A}^{\prime\nu}-\partial^{\eta}\mathrm{A}^{\prime\nu})^* \mathrm{J}_{\eta}^{\prime}/c + c.c. \Bigr) \\ \\

 \approx & \dfrac{1}{4}\Bigl( -\mathrm{i}k^{\nu}\mathrm{A}_{\alpha}^{\prime *}D^{\alpha\beta}\mathrm{A}_{\beta}^{\prime} - k^{\nu}\mathrm{A}_{\alpha}^{\prime *}\dfrac{\partial D^{\alpha\beta}}{\partial k_{\mu}}\partial_{\mu}\mathrm{A}_{\beta}^{\prime} + \partial^{\nu}\mathrm{A}_{\alpha}^{\prime *}D^{\alpha\beta}\mathrm{A}_{\beta}^{\prime} + \mathrm{A}^{\prime *\nu} \underbrace{(\mathrm{i}k^{\eta}\mathrm{J}_{\eta}^{\prime}+\partial^{\eta}\mathrm{J}_{\eta}^{\prime})}_0/c - \partial^{\eta}(\mathrm{A}^{\prime *\nu}\mathrm{J}_{\eta}^{\prime})/c + c.c. \Bigr) \\ \\

 = & - \partial_{\mu} \left( \dfrac{1}{4} \mathrm{A}_{\alpha}^{\prime *} \Bigl( k^{\nu} \dfrac{\partial D^{\alpha\beta}}{\partial k_{\mu}} - \eta^{\mu\nu} D^{\alpha\beta} + \eta^{\alpha\nu}D^{\mu\beta} + D^{\alpha\mu}\eta^{\nu\beta} \Bigr) \mathrm{A}_{\beta}^{\prime} \right).

\end{array}
\end{equation}
\end{widetext}

\noindent  Here we have only kept terms that are at most first order in the derivatives of the slowly varying 4-vector amplitude $\mathrm{A}_{\mu}^{\prime}(x)$.  In the second line of the above equation we take advantage of the law of conservation of electric charge $\mathrm{i}k_{\mu}\mathrm{J}^{\prime\mu}+\partial_{\mu}\mathrm{J}^{\prime\mu}=0$.  We have also taken advantage of the fact that the medium we are considering is lossless, resulting in the reciprocal differential operator being Hermitian.  Eq.~(\ref{f_av}) can be rewritten as 

\begin{equation}\label{conservation}
\mathrm{F}^{\nu} + \partial_{\mu}T^{\mu\nu} = 0,
\end{equation}

\noindent which is simply the time averaged law of conservation of energy and momentum where the time averaged EM Tensor $T^{\mu\nu}$ is defined as

\begin{equation}\label{EM_Tensor}
T^{\mu\nu} =  \dfrac{1}{4} \mathrm{A}_{\alpha}^{\prime *} \Bigl( k^{\nu} \dfrac{\partial D^{\alpha\beta}}{\partial k_{\mu}} - \eta^{\mu\nu} D^{\alpha\beta} + \eta^{\alpha\nu}D^{\mu\beta} + D^{\alpha\mu}\eta^{\nu\beta} \Bigr) \mathrm{A}_{\beta}^{\prime}.
\end{equation}

\noindent Eq.~(\ref{EM_Tensor}) is invariant under the gauge transformation $\mathrm{A}_{\mu}^{\prime}(x)\rightarrow\mathrm{A}_{\mu}^{\prime}(x)+\mathrm{i}k_{\mu}\lambda^{\prime}(x)$, where $\lambda^{\prime}(x)$ is a slowly varying complex valued transformation parameter.

Our EM Tensor appears somewhat analogous to the normal vacuum EM Tensor as derived from the Lagrangian density $\mathcal{L}$, or $T^{\mu\nu} = \partial\mathrm{A}_{\alpha}/\partial x_{\nu} \cdot \partial \mathcal{L}/\partial(\partial\mathrm{A}_{\alpha}/\partial x_{\mu}) - \eta^{\mu\nu}\mathcal{L}$~\cite{Landau_EM_Tensor}.  Indeed, this was the motivation for pursuing this form of the EM Tensor.  The difference between the two expressions is that we have allowed for the dependence of the constitutive parameters on $\omega$ and $\mathbf{k}$.  The price we pay for this assumption is that we have assumed the electromagnetic field to be a nearly monochromatic wave.  One similarity between our Eq.~(\ref{EM_Tensor}) and the more standard version is that both assume no losses in the medium.  This is a limitation which must be emphasized since this work was motivated by research in metamaterials and many metamaterials are very lossy.
 Finally, a similar expression for the energy momentum tensor is presented in Ref.~\onlinecite{Dewar_77}.  The difference between the two expressions are the two extra terms in Eq.~(\ref{EM_Tensor}) which ensure gauge invariance.

\section{Minkowski}

It is convenient to separate the EM Tensor into four separate parts

\footnotesize
\begin{equation}\label{TSGT}
\begin{array}{rlrl}

\mathrm{U} = & T^{00} = \dfrac{1}{4}\mathrm{A}_{\alpha}^{\prime *}\left( \omega\dfrac{\partial D^{\alpha\beta}}{\partial\omega}-D^{\alpha\beta} + \eta^{\alpha 0}D^{0\beta} + D^{\alpha 0}\eta^{0\beta} \right)\mathrm{A}_{\beta}^{\prime}, \\ \\

\mathrm{S}_i = & c T^{i0} = \dfrac{1}{4}\mathrm{A}_{\alpha}^{\prime *}\left( -\omega\dfrac{\partial D^{\alpha\beta}}{\partial k_i} + c\eta^{\alpha 0}D^{i\beta} + cD^{\alpha i}\eta^{0\beta} \right)\mathrm{A}_{\beta}^{\prime}, \\ \\

\mathrm{G}_j = & T^{0j}/c = \dfrac{1}{4}\mathrm{A}_{\alpha}^{\prime *}\left( k_j\dfrac{\partial D^{\alpha\beta}}{\partial\omega} + \dfrac{\eta^{\alpha j}}{c}D^{0\beta} + D^{\alpha 0}\dfrac{\eta^{j\beta}}{c} \right)\mathrm{A}_{\beta}^{\prime}, \\ \\

\Sigma_{ij} = & T^{ij} = \dfrac{1}{4}\mathrm{A}_{\alpha}^{\prime *}\left( -k_j\dfrac{\partial D^{\alpha\beta}}{\partial k_i}+\delta_{ij}D^{\alpha\beta} + \eta^{\alpha j}D^{i\beta} + D^{\alpha i}\eta^{j\beta} \right)\mathrm{A}_{\beta}^{\prime},

\end{array}
\end{equation}
\normalsize

\noindent  which are the energy density $\mathrm{U}$, the energy flux $\mathbf{S}$, the momentum density $\mathbf{G}$, and the momentum flux (stress tensor) $\Sigma_{ij}$.  Here we have introduced the 3 dimensional wavevector component $k_i=\mathbf{k}\cdot\hat{\mathbf{e}}_i$.

For practical purposes, we are often only interested in the EM Tensor for freely propagating waves not coupled to a source (charge-current).  For such a freely propagating wave uncoupled to a source, the frequency and wavevector are connected by a dispersion relation $\omega=\omega(\mathbf{k})$ which satisfies $\mathrm{det}(D^{\alpha\beta}(\omega(\mathbf{k}),\mathbf{k}))=0$.  This implies that zero is an eigenvalue of $D^{\alpha\beta}(\omega(\mathbf{k}),\mathbf{k})$ or $D^{\alpha\beta}(\omega(\mathbf{k}),\mathbf{k})a_{\beta}=0$ where $a_{\beta}$ is a polarization eigen 4-vector.  By taking full derivatives of this quantity with respect to the wavevector while restricting ourselves to the dispersion relation we get

\begin{equation}\label{dD}
0 = \dfrac{d}{d k_i} (D^{\alpha\beta}a_{\beta}) = \left( \dfrac{\partial D^{\alpha\beta}}{\partial k_i} + \dfrac{\partial\omega}{\partial k_i}\dfrac{\partial D^{\alpha\beta}}{\partial\omega} \right) a_{\beta}.
\end{equation}

\noindent  We have assumed that the polarization eigen 4-vector $a_{\beta}$ is independent of the frequency and wavenumber.  This is not true in general but is a necessary requirement for a pulse to propagate without deformation.  Eq.~(\ref{dD}) reveals that for freely propagating waves, where $D^{\alpha\beta}\mathrm{A}_{\beta}^{\prime}=0$, there are several useful relationships between the various components of the EM Tensor

\begin{equation}\label{relationships}
\begin{array}{rl}

\mathrm{S}_i = & \dfrac{\partial\omega}{\partial k_i}\mathrm{U}, \\ \\

\mathrm{G}_j = & \dfrac{k_j}{\omega} \mathrm{U}, \\ \\

\Sigma_{ij} = & \dfrac{\partial\omega}{\partial k_i}\mathrm{G}_j, \\ \\

\Sigma_{ij} = & \dfrac{k_j}{\omega}\mathrm{S}_i.

\end{array}
\end{equation}

\noindent  Here we see the appearance of the group velocity $\mathbf{v}_g=\hat{\mathbf{e}}_i\partial\omega/\partial k_i$, as well as the phase velocity $\mathbf{v}_{\phi}=\mathbf{k}\omega/(\mathbf{k}\cdot\mathbf{k})$.  The fact that the energy density and energy flux are related by the group velocity guarantees energy conservation in a pulse.  The analogous relationship between the momentum density and momentum flux ensures momentum conservation.  The relationship between the energy density and the momentum density involving the phase velocity conforms with our quantum mechanical understanding of wave energy and momentum~\cite{Nelson_91}.

Finally, for a non-temporally and non-spatially dispersive medium, Eq.~(\ref{EM_Tensor}) simplifies to the familiar Minkowski EM Tensor\cite{Kong_Tensor}.  Given the relationships between the various components of the EM Tensor given in Eq.~(\ref{relationships}), as well as the fact that the EM Tensor is clearly unsymmetric, Eq.~(\ref{EM_Tensor}) should be considered a more general version of the Minkowski EM Tensor.

\section{Samples}

The constitutive parameters defined in Eq.~(\ref{constitutive}) were chosen for their convenience when working with the electromagnetic 4-vector $\mathrm{A}_{\mu}$.  Still, they differ from the set of constitutive parameters that many people are probably familiar with

\begin{equation}\label{constitutive_eps}
\begin{array}{rl}

\mathbf{D}(\omega,\mathbf{k}) = & \hat{\epsilon}(\omega,\mathbf{k})\cdot\mathbf{E}(\omega,\mathbf{k}) + \hat{\xi}(\omega,\mathbf{k})\cdot\mathbf{H}(\omega,\mathbf{k}), \\ \\

\mathbf{B}(\omega,\mathbf{k}) = & \hat{\zeta}(\omega,\mathbf{k})\cdot\mathbf{E}(\omega,\mathbf{k}) + \hat{\mu}(\omega,\mathbf{k})\cdot\mathbf{H}(\omega,\mathbf{k}).

\end{array}
\end{equation}

\noindent  The relationship between the two sets of constitutive parameters is given by 

\begin{equation}
\begin{array}{rlrl}
\hat{p} = & \hat{\epsilon}-\hat{\xi}\cdot\hat{\mu}^{-1}\cdot\hat{\zeta},
\ \ \ \ &
\hat{l} = & \hat{\xi}\cdot\hat{\mu}^{-1},
\\ \\
\hat{m} = & -\hat{\mu}^{-1}\cdot\hat{\zeta},
\ \ \ \ &
\hat{q} = & \hat{\mu}^{-1}.
\end{array}
\end{equation}

\noindent Combining Eqs.~(\ref{D_oper}) and (\ref{C_con}), we can write the reciprocal differential operator as

\begin{equation}\label{Diff}
\begin{array}{rl}

D^{00} = & k_ap_{ab}k_b,
\\ \\
D^{i0} = & \left(\dfrac{\omega}{c}p_{ib}+\epsilon_{ian}k_am_{nb}\right)k_b,
\\ \\ 
D^{0j} = & k_a \left( p_{aj}\dfrac{\omega}{c}+l_{am}\epsilon_{mbj}k_b \right),
\\ \\
D^{ij} = & \dfrac{\omega^2}{c^2}p_{ij} + \dfrac{\omega}{c}l_{im}\epsilon_{mbj}k_b \\ \\
 & + \epsilon_{ian}k_am_{nj}\dfrac{\omega}{c} + \epsilon_{iam}k_aq_{mn}\epsilon_{nbj}k_b.

\end{array}
\end{equation}

\noindent  We also define the constitutive matrix

\begin{equation}\label{Con_M}
\hat{C} = 
\left(\!\!\begin{array}{cc}
\hat{\epsilon} & \hat{\xi} \\[0pt]
\hat{\zeta} & \hat{\mu}
\end{array}\!\!\right).
\end{equation}

\noindent After performing a heroic feat of algebra, using Eqs.~(\ref{TSGT},\ref{Diff},\ref{Con_M}) we obtain expressions for the energy density

\begin{equation}\label{U_C}
\mathrm{U} = \dfrac{1}{4}\left(\!\!\begin{array}{c}
\mathbf{E}^{\prime} \\[4pt]
\mathbf{H}^{\prime}
\end{array}\!\!\right)^{\dagger}\cdot
\dfrac{\partial(\omega\hat{C})}{\partial\omega}\cdot
\left(\!\!\begin{array}{c}
\mathbf{E}^{\prime} \\[4pt]
\mathbf{H}^{\prime}
\end{array}\!\!\right),
\end{equation}

\noindent the energy flux

\begin{equation}
\mathbf{S} = \dfrac{\mathrm{Re}[c\mathbf{E}^{\prime}\times\mathbf{H}^{\prime *}]}{2}
- \dfrac{\hat{\mathbf{e}}_i}{4}\left(\!\!\begin{array}{c}
\mathbf{E}^{\prime} \\[4pt]
\mathbf{H}^{\prime}
\end{array}\!\!\right)^{\dagger}\cdot
\dfrac{\partial(\omega\hat{C})}{\partial k_i}\cdot
\left(\!\!\begin{array}{c}
\mathbf{E}^{\prime} \\[4pt]
\mathbf{H}^{\prime}
\end{array}\!\!\right),
\end{equation}

\noindent the momentum density

\begin{equation}
\mathbf{G} = \dfrac{\mathrm{Re}[\mathbf{D}^{\prime}\times\mathbf{B}^{\prime *}]}{2c} - 
\dfrac{\hat{\mathbf{e}}_i}{4}\left(\!\!\begin{array}{c}
\mathbf{D}^{\prime} \\[4pt]
\mathbf{B}^{\prime}
\end{array}\!\!\right)^{\dagger}\cdot
\dfrac{\partial(k_i\hat{C}^{-1})}{\partial\omega}\cdot
\left(\!\!\begin{array}{c}
\mathbf{D}^{\prime} \\[4pt]
\mathbf{B}^{\prime}
\end{array}\!\!\right),
\end{equation}

\noindent and the momentum flux (stress tensor)

\begin{equation}\label{T_C}
\Sigma_{ij} = - \dfrac{\mathrm{Re}[\mathrm{D}_i^{\prime *}\mathrm{E}_j^{\prime}+\mathrm{B}_i^{\prime *}\mathrm{H}_j^{\prime}]}{2}
 + \dfrac{1}{4}\left(\!\!\begin{array}{c}
\mathbf{D}^{\prime} \\[4pt]
\mathbf{B}^{\prime}
\end{array}\!\!\right)^{\dagger}\cdot
\dfrac{\partial(k_j\hat{C}^{-1})}{\partial k_i}\cdot
\left(\!\!\begin{array}{c}
\mathbf{D}^{\prime} \\[4pt]
\mathbf{B}^{\prime}
\end{array}\!\!\right).
\end{equation}

\noindent  Here we have used the complex valued field amplitudes

\begin{equation}
\begin{array}{rl}
\mathbf{E}^{\prime} = & \mathrm{i}\mathbf{k}\mathrm{A}_0^{\prime}-\mathrm{i}\frac{\omega}{c}\mathbf{A}^{\prime}, \\[4pt]
\mathbf{B}^{\prime} = & -\mathrm{i}\mathbf{k}\times\mathbf{A}^{\prime}, \\[4pt]
\mathbf{D}^{\prime} = & \hat{p}\cdot\mathbf{E}^{\prime}+\hat{l}\cdot\mathbf{B}^{\prime}, \\[4pt]
\mathbf{H}^{\prime} = & \hat{m}\cdot\mathbf{E}^{\prime}+\hat{q}\cdot\mathbf{B}^{\prime}.
\end{array}
\end{equation}

The expressions for the energy density $\mathrm{U}$ and Poynting Flux $\mathbf{S}$ have been previously derived by others~\cite{Agranovich_Energy,Landau_Energy,Kamenetskii_96}.  A similar expression for the momentum density $\mathbf{G}$  in temporally dispersive dielectrics was derived in~\cite{Nelson_91}.  The full expressions for the momentum density $\mathbf{G}$ and momentum flux $\Sigma_{ij}$ have, to our knowledge, never been published, but they can easily be derived using the same arguments forwarded in Refs.~\onlinecite{Agranovich_Energy,Landau_Energy,Kamenetskii_96}

Our expression for the EM Tensor in Eq.~(\ref{EM_Tensor}) is complicated primarily because it is very general.  However, metamaterial researchers are often interested in highly symmetric crystals with waves propagating along a principle axis.  Assuming high enough symmetry, this situation can often reduce to a problem involving just four constitutive parameters.  For example, imagine a cubic metamaterial crystal with reflection symmetries in the direction of each of the principle axes, $\hat{\mathbf{e}}_x$, $\hat{\mathbf{e}}_y$ and $\hat{\mathbf{e}}_z$.  For a wave propagating in the $\hat{\mathbf{e}}_x$ direction with the electric field polarized in the $\hat{\mathbf{e}}_y$ direction, we need only consider the four constitutive parameters in the relation

\begin{equation}
\left(\!\!\begin{array}{c}
\mathrm{D}_y^{\prime} \\[4pt]
\mathrm{B}_z^{\prime} 
\end{array}\!\!\right) = 
\left(\!\!\begin{array}{cc}
\epsilon_{yy} & \xi_{yz} \\[4pt]
\zeta_{zy} & \mu_{zz}
\end{array}\!\!\right)\cdot
\left(\!\!\begin{array}{c}
\mathrm{E}_y^{\prime} \\[4pt]
\mathrm{H}_z^{\prime} 
\end{array}\!\!\right).
\end{equation}

\noindent  The reciprocal differential equation in Eq.~(\ref{wave_eq}) reduces to 

\begin{equation}
\dfrac{1}{c}\mathrm{J}_y^{\prime} = -D^{yy}(\omega,k_x)\mathrm{A}_y^{\prime},
\end{equation}

\noindent where the reciprocal differential operator is

\begin{equation}
\begin{array}{c}

D^{yy}(\omega,k_x) = p_{yy}\dfrac{\omega^2}{c^2} + \dfrac{\omega}{c}k_xl_{yz} - \dfrac{\omega}{c}k_xm_{zy} - k_x^2q_{zz} \\ \\

 = \left(\epsilon_{yy}-\dfrac{\xi_{zy}\zeta_{yz}}{\mu_{zz}}\right)\dfrac{\omega^2}{c^2} + \dfrac{\omega}{c}k_x\dfrac{\xi_{yz}}{\mu_{zz}} + \dfrac{\omega}{c}k_x\dfrac{\zeta_{zy}}{\mu_{zz}} - \dfrac{k_x^2}{\mu_{zz}},

\end{array}
\end{equation}

\noindent and the EM Tensor simplifies to 

\begin{equation}
\left(\!\!\begin{array}{cc}
\mathrm{U} & c\mathrm{G}_x \\[4pt]
\dfrac{\mathrm{S}_x}{c} & \Sigma_{xx}
\end{array}\!\!\right) = 
\left(\!\!\begin{array}{cc}
\omega\dfrac{\partial D^{yy}}{\partial\omega} - D^{yy} & ck_x\dfrac{\partial D^{yy}}{\partial\omega} \\[9pt]
-\dfrac{\omega}{c}\dfrac{\partial D^{yy}}{\partial k_x} & -k_x \dfrac{\partial D^{yy}}{\partial k_x} + D^{yy}
\end{array}\!\!\right)\dfrac{\vert\mathrm{A}_y^{\prime}\vert^2}{4}.
\end{equation}

\noindent  Again, note that the dispersion relation for such a wave is $D^{yy}(\omega,k_x) = 0$.

\section{Loss}

Throughout this paper we have consistently assumed that the medium in question was lossless. This assumption is important in every previous derivation in this paper.  Here we slightly relax this requirement and allow for a medium that has small but nonzero losses.

In Eq.~(\ref{f_av}) we see the two terms $-\mathrm{i}k^{\nu}\mathrm{A}_{\alpha}^{\prime *}D^{\alpha\beta}\mathrm{A}_{\beta}^{\prime} + c.c.$  Due to the lossless nature of the medium expressed by the relation $D^{\alpha\beta} = (D^{\beta\alpha})^*$, these two terms cancel.  If we allow for the medium to have small losses, the sum of these terms becomes nonzero, and the energy conservation equation expressed in Eq.~(\ref{conservation}) now becomes 

\begin{equation}\label{F_loss}
\mathrm{F}^{\nu} + \partial_{\mu}T^{\mu\nu} + \Gamma^{\nu} = 0,
\end{equation}

\noindent where we have introduced the loss 4-vector

\begin{equation}\label{Gamma}
\Gamma^{\nu}  = -\dfrac{k^{\nu}}{2} 
\left(\!\!\begin{array}{c}
\mathbf{E}^{\prime} \\[4pt]
\mathbf{H}^{\prime}
\end{array}\!\!\right)^{\dagger}\cdot
\left(\dfrac{\hat{C}-\hat{C}^{\dagger}}{2\mathrm{i}}\right) \cdot
\left(\!\!\begin{array}{c}
\mathbf{E}^{\prime} \\[4pt]
\mathbf{H}^{\prime}
\end{array}\!\!\right).
\end{equation}

\noindent  It should be noted that the sign of Eq.~(\ref{Gamma}) depends on the fact that we used the $e^{\mathrm{i}k\cdot x} = e^{\mathrm{i}(\omega t-\mathbf{k}\cdot\mathbf{x})}$ wave convention.  $\Gamma^{\nu}$ is a force density 4-vector.  The $\Gamma^0$ term should be interpreted as the power loss per unit volume (divided by a factor of $c$).  The $\Gamma^i$ terms should be interpreted as a force per unit volume acting on the medium.

All of the expressions that we have derived, the EM Tensor and its decomposition into separate components, are still approximately correct as long as the loss is small.  However, metamaterials with metallic inclusions often have very large losses.  All of the expressions in this paper fail when applied to materials with sufficiently large losses. \\

A few final comments.  We have not addressed the periodic nature of a crystal.  Early in this paper we assumed that our medium has infinitesimal translational symmetry.  This is not true for a crystal, which only has discrete translational symmetry.  However, though we do not show this here, a similar analysis can be performed with eigenmodes of a periodic crystal of the nature $\slashed{D}u=\lambda u$ where $\slashed{D}$ is a differential operator describing the microscopic geometry and constitutive properties of the crystal, $\lambda$ is an eigenvalue, $u$ is a potential field eigenmode, and $\lambda u$ represents a source.  With a consistent normalization, $\mathrm{U} \propto \omega\cdot\partial(\lambda u^{\dagger}\cdot u)/\partial\omega$ replaces $\mathrm{U} \propto \omega\cdot\partial D/\partial\omega$, with similar substitutions for the other EM Tensor components.  It should be noted however that in a crystal both the momentum density $\mathbf{G}$ and the momentum flux $\Sigma_{ij}$ have a $2\pi/a_i$ ambiguity associated with the wavevector $k_i$.  Addressing this issue is beyond the scope of this paper.

It is sometimes convenient to solve electromagnetic problems using the dual potential fields consisting of $\mathrm{C}_0$, the magnetoelectric pseudoscalar potential, and $\mathbf{C}$, the magnetoelectric pseudovector potential.  The procedure for deriving the EM tensor using the magnetoelectric 4-pseudovector is similar to the one demonstrated here.  Eqs.~(\ref{U_C}-\ref{T_C}) remained unchanged.

Finally, this is merely speculation, but the usefulness of the reciprocal differential operator $D^{\alpha\beta}(k)$ in this paper suggests that it might provide a pathway towards progress on the longstanding problem of metamaterial homogenization.

\end{document}